\documentclass[prd,twocolumn,nofootinbib]{revtex4}
\usepackage{amsmath,amssymb}
\usepackage{graphicx}
\usepackage{rotating}
\usepackage{bm}

\usepackage{color}

\newcommand{\ee}{\begin{equation}}
  \newcommand{\eee}{\end{equation}}
\newcommand{\ea}{\begin{eqnarray}}
  \newcommand{\eea}{\end{eqnarray}}

\definecolor{unterlegung}{rgb}{0.92,0.92,0.92}
\newlength{\importantlength}
\newcommand{\mpc}{\textrm{Mpc}}

\newcommand{\cfa}{\textsc{cmbfast}}
\newcommand{\ce}{\textsc{cmbeasy}}
\newcommand{\cmbfast}{\cfa}
\newcommand{\cmbeasy}{\ce}

\newcommand{\ome}[2][]{\ensuremath{
    \Omega_{#1}^{#2}}}

\newcommand{\omq}{\ome[0]{\varphi}}

\newcommand{\omc}{\ome[0]{\rm c}}
\newcommand{\omb}{\ome[0]{\rm b}}

\newcommand{\class}[1]{{\tt #1}}

\begin{document}

\author{Michael Doran}
\email{Michael.Doran@Dartmouth.edu}
\url{www.cmbeasy.org}
\affiliation{Department of Physics \& Astronomy,
        HB 6127 Wilder Laboratory,
        Dartmouth College,
        Hanover, NH 03755, USA}
\affiliation{Institut f\"ur Theoretische Physik der Universit\"at Heidelberg, Philosophenweg 16, 69120 Heidelberg, Germany}

\title{CMBEASY:: an Object Oriented Code for the Cosmic Microwave Background}

\begin{abstract}
We have ported the \cmbfast\ package to the C++ programming language
to produce \mbox{\ce}, an object oriented code for the cosmic microwave
background. The code is available at www.cmbeasy.org. 
We sketch the design of the new code, emphasizing the benefits
of object orientation in cosmology, which allow for simple
substitution of different cosmological models and gauges. Both gauge invariant
perturbations and quintessence support has been added to the code. For
ease of use, as well as for instruction, a graphical user interface is
available.
\end{abstract}

\maketitle

\section{Introduction}\label{sec::cmbeasy}
\setlength{\importantlength}{\columnwidth}
\addtolength{\importantlength}{-0.75em}

The \cmbfast\ computer program for calculating cosmic microwave
background (CMB) temperature and polarization anisotropy spectra,
implementing the fast line-of-sight integration method, has been
publicly available since 1996 \cite{Seljak:1996is}. It has been widely used to calculate
spectra for open, closed, and flat cosmological models containing
baryons and photons, cold dark matter, massless and massive neutrinos,
and a cosmological constant. It is a very well-tested program that has
enabled many cosmologists to compare models of the Universe against the
experimental CMB data.
                
However, from the point of view of code design, there are few programs
that could not be improved. This is also true for \cmbfast: it is a
rather monolithic code that is quite difficult to oversee and modify.
For example, variations in the cosmic expansion history due to dark
energy necessitate similar changes to the code in numerous separate
locations.

In order to address these shortcomings and simplify modifications
of the code -- in our case the implementation of quintessence
models and gauge invariant variables -- we have ported the 
\cmbfast\ package to the C++ programming language. 
The C++ language is \emph{object} oriented and it turns out that
to think in objects (more of this soon), is very advantageous
in cosmology. 

The program has not been rewritten from scratch,
but redesigned step by step. Some people may argue that it is
hence not \emph{independent}, i.e. some unknown errors and limitations
in \cmbfast\ could be present in the new code. The object oriented
modular design, however, ensures that each part of the code
is independently testable. If, for instance, one does not trust the integrator,
one can use another one to check it, without changing anything
else in the package. In addition most of the lines in the code
\emph{have} been rewritten, to benefit from the redesign.

\cmbeasy\ calculates the temperature and polarization anisotropies for
spatially flat universes, containing dark energy, dark matter, baryons
and neutrinos. The dark energy can be a cosmological constant  or quintessence.
There are three scalar field field models implemented, or alternatively one
may specify an equation-of-state history. Furthermore, the modular code is
easily adapted to new problems in cosmology.

In the next section \ref{sec::objects} we explain object oriented programming and
its use in cosmology. The design of \ce\ is presented in section \ref{sec::design},
while the graphical user interface is discussed in section \ref{sec::gui}. Computing
precision is discussed in  section \ref{sec::prec}, while 
the documentation for \ce\ is  introduced in \ref{sec::documentation}. For an introduction
to perturbation theory and CMB physics using the conventions of  \ce,
as well as a summary of the evolution equations, please see the documentation on www.cmbeasy.org.

\section{The code}
We briefly introduce the concept of object oriented programming before turning
to the design of \ce. We do this using examples from the code.

\subsection{Objects}\label{sec::objects}
In C++, a class is a user-defined data type \cite{stroustrup}. 
As with any data type, there can be many variables with
the same type, e.g. floats, arrays, complex numbers.  In the case of user-defined types, 
the specific variables are called objects. 

\subsubsection{Encapsulation}
Quite often, some data and functions acting on the data 
are so tightly connected, that it is sensible to think of
them as one object. As an example, let us discuss splines.
Given a discrete set of $n$ points $x_i$ with $x_i < x_{i+1}$ and 
corresponding $f(x_i)=y_i$,  a spline can smoothly interpolate,
i.e. give $f(x)$ for any $x \in [x_0,x_n]$. Provided 
that the sampling is dense enough, arbitrary functions may be 
described by a spline for all practical purposes. This is widely
used in \cmbfast. For instance, the $C_l$'s are calculated only every
$50$ l-values for $l > 200$. As the spectrum is very smooth, this
still gives a precise result. 

\begin{figure}[!t]
\begin{center}
\includegraphics[angle=0,scale=0.33]{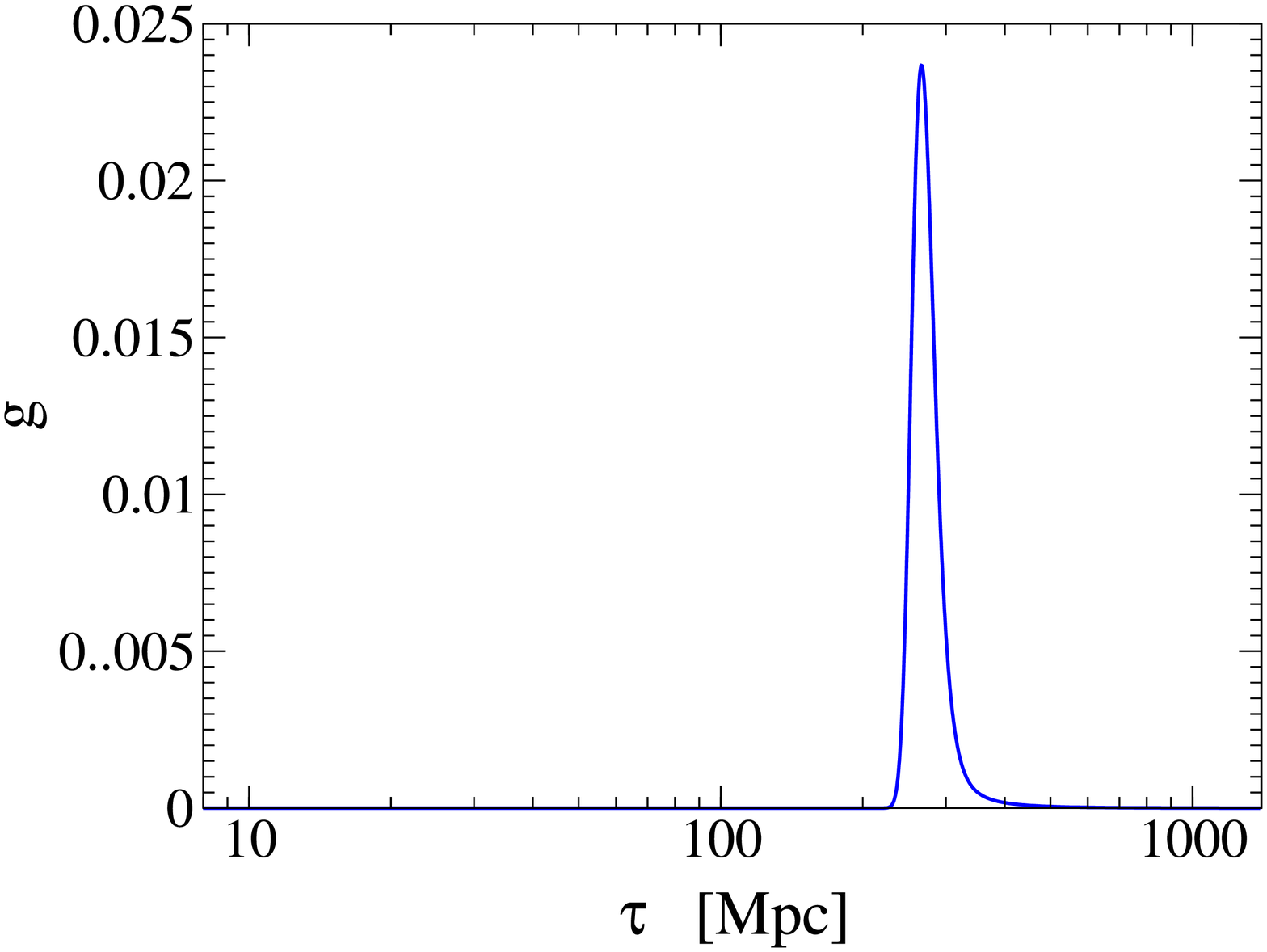}
\caption{\label{fig::visibility}The visibility $g \equiv \dot \kappa \exp(\kappa(\tau)-\kappa(\tau_0))$ 
as a function of conformal time $\tau$ in $\mpc$. Its peak at about
$\tau \approx 300\,\mpc$ defines the epoch of last scattering.
Before the visibility function peaks, photons are very likely to scatter again until
the Universe becomes translucent. After the peak, photons do not scatter
at a substantial rate. It is thus the balance between frequent scattering and 
sufficiently low optical depth that will give the largest contribution towards
the anisotropy today. And this fact is exactly encoded in $g$.}
\end{center}
\end{figure}

Now, the function describing the photon visibility (see also Figure \ref{fig::visibility}), calculated in the
part of the code that tracks the thermal history of the Universe, can be
used to define a spline.
Without object orientation, one would need to keep track
of various variables, most notably arrays for the $x,y$ data and
derivatives needed for spline interpolation. Also, in order
to assure quick access within the spline data table, one either
needs to know the precise layout of the data arrays (\cmbfast\ does
this), or even more variables (storing for instance the last
interpolation $x$ value) would be necessary. In total, this
sums up to a lot of bookkeeping for a conceptually simple 
entity like a spline.  

Alternatively, one may define a \emph{class} holding all the necessary
variables a spline needs, together with \emph{definitions} of an
interface with which other parts of the program can access  and 
manipulate the spline data. An object behaves as described by the
corresponding class. There can be an arbitrary number of
objects of a  certain class (just like there is one floating point
type {\tt float}, but many variables of \emph{type} {\tt float} in a program).\footnote{
We usually denote here (and in the code) classes with capital
first letter. In some cases where there is only one object of a 
class used in the code, we denote the object with the same name
as the class, but with lower case initial letter. Hence, the line \\[1ex]
{\tt Cosmos} {\bf cosmos};
\\[1ex]
creates an object `cosmos' of the class `Cosmos'.}
 The class (in our case) called \class{Spline},
can hence be viewed as yet another data type, with no more bookkeeping
needed than say for a floating point number.
To illustrate this, let us discuss the visibility function 
$g \equiv \dot \kappa \exp(\kappa(\tau)-\kappa(\tau_0))$, where
$\kappa(\tau)$ is the optical depth.
Its typical shape is depicted in Figure \ref{fig::visibility}, and its peak defines the epoch of last scattering. 
As soon as the Spline called {\bf visibility} has been given the data, its
maximum can be determined by a single line of code:\\[2ex]
{\tt 
\indent tau\_ls = {\bf visibility}.maximum();
\\[2ex] }
Here, the function maximum() acts on the visibility.
The important point to notice is that \emph{all} functions
defined in the Spline class are immediately available to everyone who uses Splines. So, 
whenever one needs to find the maximum, integrate a spline, calculate the convolution of 
two Splines etc, this can be done using very few lines of code: the functionality is fully encapsulated
in the implementation of the Spline class. Any increase in performance or sophistication of
the Spline class immediately translates over to all Splines used in the program.

\subsubsection{Inheritance}
Tightly connected to the fact that data and methods are combined
within one object, is the concept of \emph{inheritance}, which  proves very
powerful in cosmology. A class can inherit from another class (in this
context called base class).  All variables and the full functionality
that the base class implements are instantly available to the inheriting
class,\footnote{This is as if a child was born with the whole
  knowledge of its parents. No training and learning would be
  necessary.  It could instantly go and increase its capabilities
  starting from the level of its parents.} called sub-class.  The
sub-class can then re-implement functions of the base class to provide
a different functionality, or add new functions and variables. The
important point to note is that \emph{all} classes deriving from the
same base class necessarily need to provide \emph{all} the functions the
base class provides. Hence, whenever the code uses
the base class, one can substitute an  inheriting class for
the base class by simply changing one line of code.

As an illustration, let us look at
the \class{Perturbation} class of \cmbeasy\ which is designed to evolve
the perturbation equations for one $k$-mode through conformal time and
calculate the anisotropy sources. The \class{Perturbation} class defines functions to do
these calculations which other parts of the program can be sure to find implemented
in all sub-classes.  In practice, there are four classes that
inherit from it, for perturbations in gauge-invariant variables and in
synchronous gauge both with and without quintessence (see also Figure
\ref{fig::design}).  From the point of view of the rest of the
program, all of them are equally well suited.\footnote{Except for the
  fact that if one wants quintessence, the perturbation class should
  of course support it.}
Hence, when given a \class{Perturbation} object, one may e.g. ask
it to initialize the perturbation variables:
{\tt\\[1ex] 
\indent{\bf perturbation}->initialScalarPerturbations();
\\[1ex]
}
Yet, depending on the specific
sub-class the object belongs to \footnote{Representing the gauge choice and
quintessence support}, the implementation
of initialScalarPerturbations() and hence the code executed will be different.
Technically this is called polymorphism.

\subsection{Design}\label{sec::design}
\begin{sidewaysfigure}
\begin{center}
\includegraphics[scale=0.6]{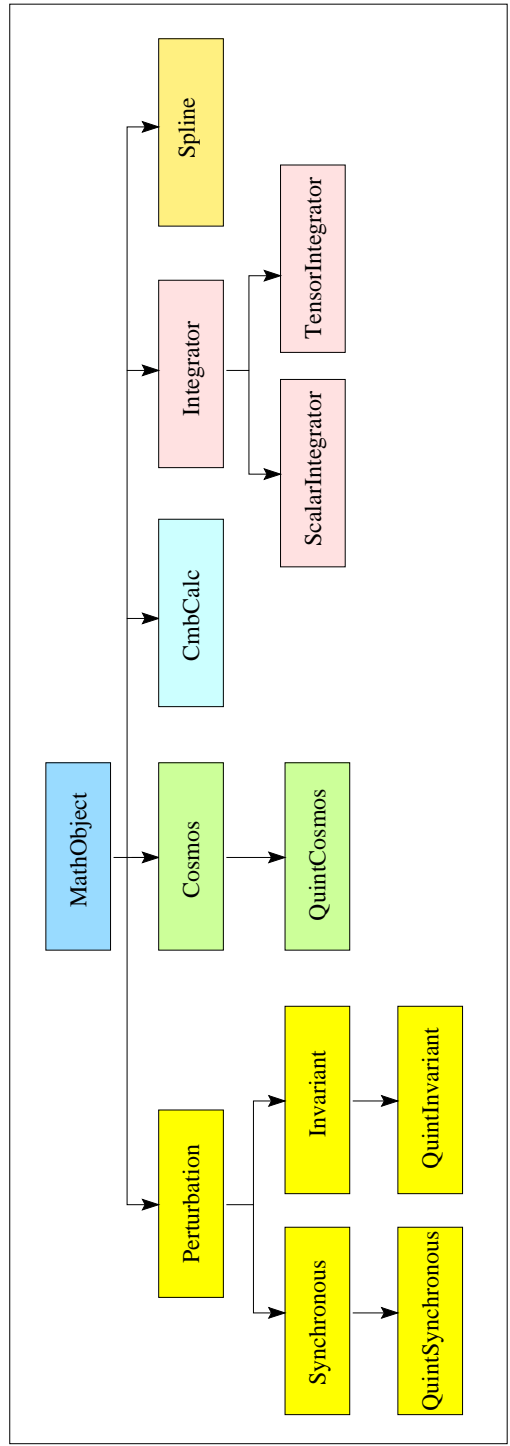}
\caption{Hierarchy of the main classes of the \cmbeasy\ package. All classes dealing
with mathematics inherit from MathObject for technical reasons. The Cosmos class
calculates the background evolution and can be extended using subclasses such
as QuintCosmos for quintessence. The perturbation equations are encapsulated
in the Perturbation class. Implementing different gauges as subclasses is therefore unproblematic.
The central instance invoking Cosmos, the Perturbations and Integrators is the 
CmbCalc class.
Not shown are several (sometimes small) classes, e.g. the ControlPanel,
which holds commonly used settings, or e.g. the MiscMath class providing
low-level mathematical functions.
}\label{fig::design}
\end{center}
\end{sidewaysfigure}

There are roughly three main steps needed to calculate the CMB 
anisotropy spectrum within the line of sight strategy:
(1) solve the expansion and thermal background evolution,
(2) propagate the perturbation equations in Fourier space,
(3) map the temperature anisotropy onto the sky today. 
In each step, several classes work together. To illustrate the
hierarchy of the most important classes, an overview is given in
Figure \ref{fig::design}. 
The core part of the package is
the \class{CmbCalc} class. It is the scheduler calling other
objects to complete all three steps. 
So let us  turn to  step (1), the expansion 
history of the universe.

\subsubsection{The background evolution}
The \class{Cosmos} class
is the central instance providing 
background quantities like the energy density $\bar\rho(\tau)$ of all species etc.
This centralization  of the background evolution within the \class{Cosmos}
class facilitates the modification of the code significantly. A different background
cosmology (such as quintessence) can be implemented by just inheriting from
\class{Cosmos} and re-implementing the expansion history part of the code.\footnote{All in all 800
 lines of a total of 2500 lines of \class{Cosmos}.}
For most other objects in the code, it makes no difference \emph{exactly what} caused the 
universe to expand, as long as they can rely on \class{Cosmos} or any sub-class to
provide access to the expansion history. Both the expansion and the thermal history are 
computed at the beginning of the calculation and stored in splines. 

\subsubsection{Propagation in k-space}
All fluctuation equations are handled by \class{Perturbation} and its sub-classes.
Its most prominent function is {\tt propagateScalar($\tau_{1},\, \tau_{2}$) } which evolves 
the perturbations from conformal time $\tau_1$ to $\tau_2$. For numerical
stability, one usually treats the era of tight coupling between  baryons and photons 
differently. However, this results in a non-analytic jump in the r.h.s of the perturbation
equations at $\tau_{tc}$. Some integrators for ordinary differential equations get confused
by this. Therefore,  {\tt propagateScalar()} will determine if the jump occurs
within [$\tau_{1},\, \tau_{2}$] and if so, integrate up to $\tau_{tc}$, make a tiny\footnote{More specific, it adds $\Delta \tau$, 
such that for double precision variables $\tau_{tc} + \frac{1}{2} \Delta \tau = \tau_{tc}$.}  
jump and integrate on till $\tau_{2}$.
At the end of each propagation, the sources of cmb anisotropies 
are calculated by the  {\tt scalarSources()} function.

\subsubsection{The Cl's}
The convolution of the sources with the Bessel functions,
as well as the final $k$-integration  are performed in
the \class{ScalarIntegrator} sub-class. In contrast to \cmbfast, we formulate
the convolution with the $j_l's$ as an ordinary differential equation (ODE). In fact,
the \class{Spline} class provides the convolution functionality. If higher
precision in this step is wanted, one may simply increase the desired
precision for Spline, which will pass this through to the integrator of ODE's.
As far as the $k$-integration is concerned, we stick to the interpolation
points of \cmbfast. We have written two other implementations
of \class{ScalarIntegrator}, although none could compete with \cmbfast's speed.
Furthermore,
no gain in accuracy justifying the speed tradeoff was found. As far 
as the $k$-integration is concerned, a fast and independent realization 
would be desirable. With the high level features of C++, algorithms that
seem hard to implement in other languages may be feasible. 

\subsubsection{Little helpers}
The most valuable class assisting the computation may be \class{Spline}.
As splines are used all over the code, and quite often in time-critical
environments, the main design goal has been speed. In order to 
perform a spline interpolation at $x$, one needs to find the interval $x \in [x_i, x_{i+1}]$.
While \cmbfast\  hard-wires the size and layout of the $x_i$-data, this has not been an
option for a versatile spline implementation. Yet, in all
time critical situations within our calculation, two facts come to the
rescue: Firstly, interpolations occur in order and close to the last interpolation.
Hence, the next value of $x$ is likely to lie within the same interval (or close to it).
 Secondly,
many splines with exactly the same $x$-data are evaluated at the same $x$. For instance, the 
r.h.s of the perturbation equations need background quantities all at the same
$\tau$. As all background quantities get their data at the same $\tau_i$ in \class{Cosmos},
they necessarily share the same $x$-data.
Therefore, some parts of the interpolation formulae have to be
computed only once\footnote{With ``once'', we mean until interpolation is requested
for a different $\tau$.}  for each $\tau$. In addition, finding the position of
the interpolation interval is simple: it will be the same for the same $\tau$.

To accommodate for these groups of splines and caching, splines come in two
flavors: with and without its own $x$-data. While a spline with its own $x$-data 
can live on its own, a spline without $x$-data can only be created 
if a ``mother'' spline providing the $x$-data is given upon  the creation of the ``child'' spline.
Within these ``families'' of shared $x$-data, interpolation
at the same $x$ is particularly fast.

Not only must a spline be fast, it must also be robust.
Hence,
a spline will allocate more memory if needed\footnote{and free unused memory},
check that no access to data occurs outside its bound, check that 
all data is ordered $x_i < x_{i+1}$ and make sure that the interpolation
tables have been built before any interpolation takes place.
Splines can easily be visualized. To write the data\footnote{and 
interpolated values, if desired} to a file, simply call \\[2ex]
{\tt \indent {\bf myspline}.dump("filename")} \\[2ex]
The anisotropy sources,
are evaluated on a $\{k, \tau\}$ grid. For the final $\tau$ and $k$ integration,
one needs to interpolate at intermediate points. This is conveniently
done using the class \class{SplineWeb}. It is essentially an array of
Splines, one for each $k$ and one for each $\tau$ value. If one 
needs to interpolate at a given $k$ value, one can ask \class{SplineWeb}
to return a \class{Spline} with x-data in the $\tau$ direction:\\[2ex]
{\tt
\indent {\bf spline\_at\_k} = {\bf web}.createAlongY(k);
\\[2ex] }
This is used to get the splines which are convoluted with the
Bessel functions. It is also very convenient for making slices through
the cold dark matter power spectrum (at either fixed $k$ or $\tau$).
Given for instance the cold dark matter \class{SplineWeb} {\bf power}, one can
dump it to a file:\\[2ex]
{\tt
\indent {\bf power}.dumpAlongX($\tau$,"filename");
\\[2ex]}
The above command 
would save the spectrum at arbitrary conformal
time $\tau$ to disk.

There are some more of these ``helpers'', most of them small classes
that sometimes make life much easier. The last one we will introduce here
 is \class{Anchor}
which keeps track of objects that are dynamically created (as are many
splines). The objects it is asked to keep track of are  deleted by it on demand, or when the \class{Anchor} object
itself is deleted. Therefore, it greatly simplifies memory handling and prevents
memory leaks. 

\subsubsection{Quintessence Implementation}
The different background evolution of quintessence \cite{Wetterich:1988fm, Ratra:1988rm,Caldwell:1998ii}
scenarios
is implemented using the \class{QuintCosmos} and the
 \class{Quintessence} class. Each subclass
of \class{Quintessence} corresponds to a certain model, such
as the exponential potential \cite{Wetterich:1988fm, Peebles:1988ek, Ratra:1988rm}, inverse power law \cite{Ratra:1988rm},  leaping kinetic term \cite{Hebecker:2001zb} etc. Certainly, a more monolithic design 
with the quintessence models implemented in the \class{QuintCosmos}
class would have been possible. However, we believe that
the \emph{details} of a model are best kept to a class of its own.
For instance tuning model parameters in order to get the right amount of
$\omq$ is different for each model and a monolithic design
would have to call \emph{differently named} functions for \emph{different} models.
Using sub-classing, \class{QuintCosmos} (and \class{Perturbation}) 
always calls  functions with the same argument and name for all models.
 Yet, as the object
implementing the function differs for each model, the code executed
by calling the function can be totally different. 
Thus, a new quintessence model can be implemented by simply sub-classing
\class{Quintessence} and implementing a minimal set of functions, such as one 
for the scalar field potential.

\begin{figure}[!t]
\begin{center}
\includegraphics[scale=0.29]{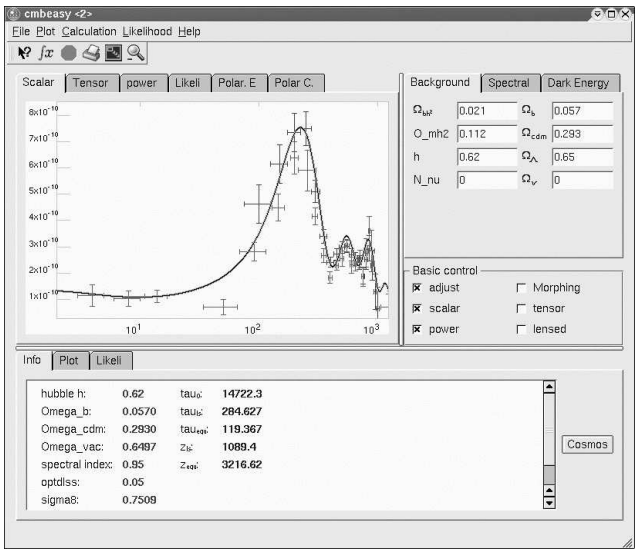}
\caption{Graphical user interface (GUI) for \cmbeasy.}\label{fig::gui}
\end{center}
\end{figure}

\subsection{Graphical User Interface}\label{sec::gui}
For educational purposes and also to simplify the parameter
input and subsequent visualization of results, a graphical user
interface (GUI) is of great value. Luckily, there is the very sophisticated 
and publicly available `Qt' library \cite{troll}   with which the 
 creation of a GUI is facilitated. Its object oriented C++ design is
a perfect match for the \cmbeasy\ package. There is therefore 
an executable program called `cmbeasy' giving interactive 
access to almost the full capabilities of the package, including quintessence.
As is seen from Figure \ref{fig::gui}, the spectra are visualized in
separate plots arranged in a so called Tab-Widget.\footnote{A widget is a part of 
a user interface that can interact. Examples are buttons, sliders, etc.}
One can for instance zoom in, select and save curves or print the plot.

\subsection{Computing precision}\label{sec::prec}
\begin{figure}[!t]
\begin{center}
\includegraphics[angle=0,scale=0.21]{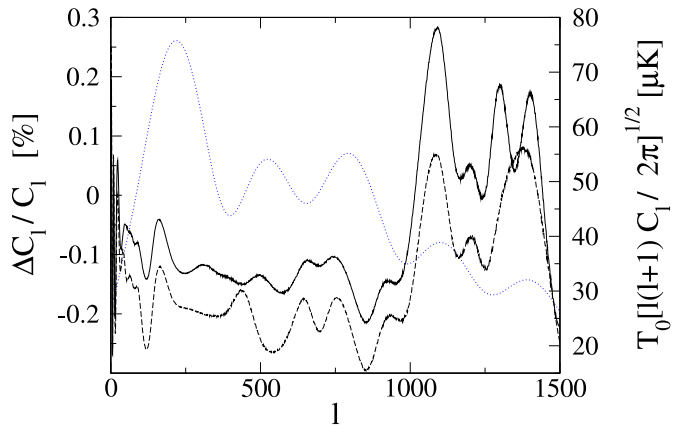}
\caption{Temperature anisotropy spectrum for 
  $h=0.65,\ \ome[0]{\Lambda} =0.6,\ \omb h^2 =0.02,\ \omc = 1 - \ome[0]{\Lambda} - \ome[0]{\rm b}$
  obtained from \cmbfast.
  The relative deviation $\Delta C_l / C_l$ of \cmbeasy's synchronous
  (long dashed line) and gauge invariant (solid line) solution with
  respect to the original \cmbfast\ spectrum are also given. 
  The accordance of all
  spectra is always better than $1\%$.  In the gauge invariant case,
  both the background and perturbation evolution as well as the $C_l$
  integration are entirely independent of the \cmbfast\ code. However, 
  they use the same thermal history algorithm that should in
  principle be independently implemented for cross checks.}\label{fig::precission}
\end{center}
\end{figure}

Comparing the results from  \cmbfast\ to those obtained from \cmbeasy, 
one generally notes accordance to better than $0.5\%$ (see also
Figure \ref{fig::precission}).

In principle, the line of sight algorithm is equivalent to evolving a
large hierarchy of multipoles.
In practice, as one truncates the hierarchy in the line of sight approach at some 
low $l$, there is a reflection of power after the higher multipoles gradually build up.
However, as by far the largest contribution towards the anisotropy occurs
around recombination (when the higher multipoles are still small), this does not
spoil the result dramatically. Yet, to check this, one may increase the number
of multipoles involved. This is true both for \cmbfast\ and \ce.

Within \cmbeasy, there are five instances in which the precision may be
increased. In three of the five cases, changing one
variable suffices. The first case is the background evolution. Here one may
want to increase (or decrease) the precision requested from the ODE integrator.
During perturbation evolution in $k$-space, the precision requirement given to
\class{Perturbation} can be changed in a similar manner. 
The same is true for the convolution of the sources with the Bessel functions, 
taking place in \class{ScalarIntegrator}. Finally,
and a bit more difficult, one may change the number and position of 
$k$-values contributing towards the $k$-integration.
This can be done by modifying \class{Integrator}, the base class of \class{ScalarIntegrator}.
Alternatively, one could devise a different implementation for the $k$-integration\footnote{i.e. a new
sub-class of \class{Integrator}}.
Finally, one may increase the number of $l$'s at which the $C_l$'s are calculated.

\subsection{Documentation}\label{sec::documentation}
Using the {\sc doxygen} program, the documentation
is automatically generated from the source code of
the \cmbeasy\ package. In its {\sc html} version, it is
interactively navigable and includes the full source code.
Due to the automatic generation, the documentation and
the code are naturally synchronized. Instructions for 
setting up and running the program are available in this
{\sc html} documentation.  
In addition, a postscript version
of the documentation is generated. 

\section{Conclusions}

We have devised an object oriented code for calculating cosmic microwave
background anisotropies. The main benefits of this code derive from its
modularity. 

The modular design allows one to test and refine each part of \ce\ 
individually. Different modules, for different numerical methods and
theoretical approaches, can easily be switched. This feature should prove
extremely useful in the development and refinement of the code. As
cosmological theory reaches for a new level of precision to match the new
generation of experiments, {\sc map} \cite{MAP}, {\sc planck} \cite{PLANCK} and other experiments will soon put
our predictions and hypotheses to the test, so future work will seek to
improve the accuracy of the code. (At the present time, \ce\  and the
standard \cmbfast\ are generally in agreement to better than $0.5\%$.)

The modular design allows one to more easily develop and build upon
\ce, to implement new cosmological models and include more physics. As
our theoretical cosmological models are a work-in-progress ---
as CMB data comes in
--- so are our tools for modeling the Universe. 
And so it makes sense to use a program design which makes
this development easy.
 Hence, the functional blocks used in
\ce\  considerably reduce the time to implement new cosmological models.
For most purposes, a new model can be achieved by sub-classing from
existing solutions, thus eliminating the need to rewrite large parts of
the code. Furthermore, new modules can be easily introduced to carry out
calculations that rely on the linear perturbation spectrum. Examples
include lensing, secondary sources in the CMB, reionization, and
cross-correlation with other tracers of large scale structure.

An additional benefit of the \ce\  package is the graphical front-end.
As we have described, the GUI allows a user to visually dial through a
range of parameters. We expect that this functionality will be useful for
developing understanding and intuition for CMB physics and cosmological
models.

As stated at the outset, there is, perhaps, no program design that could
not be improved. However, we hope that \ce\ will serve the cosmology community
well.

\begin{acknowledgments}
I would like to thank Luca Amendola and Ruth Durrer for their help, 
Uro\~s Seljak and Matias Zaldarriaga for
\cmbfast\ and their permission to distribute \cmbeasy\ under its
new name.  I would like to thank R. R. Caldwell, Martin Kunz, Christian M\"uller,
Gregor Sch\"afer and C. Wetterich for useful discussions.
Finally I would like to thank the free software
community for their excellent compilers, libraries and desktop environments.
This work was supported by NSF grant PHY-0099543 at Dartmouth and PHY-9907949 
at the KITP and the Graduiertenkolleg des Instituts f\"ur Theoretische Physik der Universit\"at Heidelberg.
\end{acknowledgments}

%=====================
%====================
%====================
%=====================
%=====================

\end{document}